\documentclass[
twocolumn,10pt,
 longbibliography,
 amsmath,amssymb,
 aps,
 pre,
]{revtex4}

\usepackage{mathtools}
\usepackage{lipsum}
\usepackage{graphicx}
\usepackage{bm}
\usepackage{comment}
\usepackage{afterpage}
\usepackage{xcolor}
\definecolor{sapphire}{HTML}{0067A5}
\usepackage[colorlinks=true,citecolor=sapphire,linkcolor=sapphire,urlcolor=sapphire]{hyperref}

\begin{document}
\title{The Jammed Phase of Infinitely Persistent Active Matter} 
\author{M. C. Gandikota$^{1}$, Rituparno Mandal$^2$, Pinaki Chaudhuri$^3$, Bulbul Chakraborty$^4$, Chandan Dasgupta$^{1,5}$}
\email{cdgupta@iisc.ac.in}
\affiliation{$^1$ International Centre for Theoretical Sciences, Tata Institute of Fundamental Research, Bengaluru 560089, India\\
$^2$ Soft Condensed Matter Group, Raman Research Institute, Bengaluru 560080, Karnataka, India\\
$^3$ The Institute of Mathematical Sciences, Taramani, Chennai 600113, India\\
$^4$ Martin Fisher School of Physics, Brandeis University, Waltham, Massachusetts 02454, USA\\ 
$^5$ Department of Physics, Indian Institute of Science, Bangalore 560012, India}

\begin{abstract}
We study an extreme active matter system, which is essentially a dense assembly of athermal, soft and infinitely persistent active particles. Using extensive numerical simulations we obtain jammed configurations of this system in two dimensions and probe the stability of such structures under increasing active forcing magnitude. We show that the critical active forcing magnitude for the jammed phase to yield scales with virial pressure as 
$f_c\sim p^\alpha$, with $\alpha=1.17$, describing the yielding line. Using a Laplacian framework, we redistribute the active forces into a modified contact force network. By analysing the statistics of these  redistributed forces, we obtain a very robust scaling law consistent with the passive limit, not just near the unjamming line, but in the entire jammed active phase. The probability distribution of the magnitude of the contact force deviates from the power-law form found in passive systems for values smaller than the active force.  Moreover, within the jammed phase, the system displays elastic, plastic, and yielding events with increasing active forcing. This active plasticity appears abruptly and can not be captured by the continuous softening of the Hessian spectrum. However, we demonstrate that the Hessian still retains the ability to predict relaxation times. These results clarify how activity modifies force distributions and leads to deformation, plasticity and yielding in dense, jammed, infinitely persistent active matter.
\end{abstract}

\maketitle

\section{Introduction}
Dense active matter~\cite{chaudhuri2021dense} occurs in diverse contexts of  biological systems such as bacterial  assembly~\cite{wensink2012meso}, epithelial tissues~\cite{angelini2011glass,henkes2020} and human crowds~\cite{gu2025emergence}.
These systems can be modelled as dense assemblies of self-propelled (active) Brownian particles that give rise to a rich phase diagram, with the persistence time of the direction of the self-propulsion force ($\tau_p$) and the magnitude of the self-propulsion force ($f_0$) serving as the primary control parameters~\cite{mandal2020extreme}. At infinite persistence time such systems remain jammed as long as the active forcing magnitude $f_0<f_c(\infty)$, where $f_c(\infty)$ is the critical active forcing needed to fluidize the system. Though many different aspects~\cite{henkes2011,ni2013pushing,berthier2013non,mandal2016active,nandi2018random,janssen2019active,tjhung2020analogies,mandal2021study,ghaznavi2025yielding} of dense active matter systems have drawn a lot of interest in recent years the detailed understanding of such actively jammed states remains mostly unaddressed.

In general, a jammed disordered solid can be unjammed in several independent ways~\cite{liu1998jamming}.  First, by reducing the packing fraction, which reduces the number of geometric constraints. Second, in the presence of thermal noise, the jammed state becomes a glass and with further increase in temperature it becomes a fluid~\cite{ikeda2013disentangling}. Third, by applying an external stress at the boundaries via shear or other mechanical deformations~\cite{heussinger2010fluctuations}. These three ways of unjamming are not exhaustive. Friction between particles and activity of the particles are known to decrease~\cite{silbert2010} and increase~\cite{briand2016,ni2013pushing,berthier2014} the critical unjamming density respectively. While with sufficient strength, active forces unjam  a dense packing of particles~\cite{liao2018criticality,mandal2020extreme,goswami2025}, how exactly it takes place and the fundamental differences of such a fluidization from the stress induced unjamming is still far from being understood.

We know that the glassy dynamics observed at finite temperature, carries the signature of underlying force-balanced jammed state of the athermal ($T=0$) limit. For example passive glasses have excess low-frequency modes (boson peak) akin to the zero temperature jammed systems. We therefore speculate that one could get insights into active glasses~\cite{berthier2013non,janssen2019active,nandi2018random, yoshida2024} when seen in the light of athermal jammed active systems. The numerical investigations of athermal active systems too indicate that the criticality of the unjamming point is carried over to both the yielding of the jammed phase and the response function in the unjammed phase~\cite{reichhardt2014,liao2018criticality}. 
Thus, a more complete understanding of the statistics of athermal jammed active states is imperative.

We examine whether jammed active systems in the limit of infinite persistence behave similarly, or fundamentally differ from, passive jammed systems. Using numerical simulations of soft (harmonic) active particles with fixed propulsion directions, the main questions we address in this paper are the following: 1) What amount of active forcing is needed to yield the jammed phase? 2) Force distributions of passive jammed systems have a power-law behaviour near the unjamming point. How does activity modify these distributions? 
3) Can we construct a Hessian of the effective potential of the system? How does the relaxation dynamics of the system relate to the Hessian? Can the Hessian predict plastic instabilities in the solid phase?

\begin{figure*}[t]
    \centering
    \includegraphics[width=0.95\textwidth]{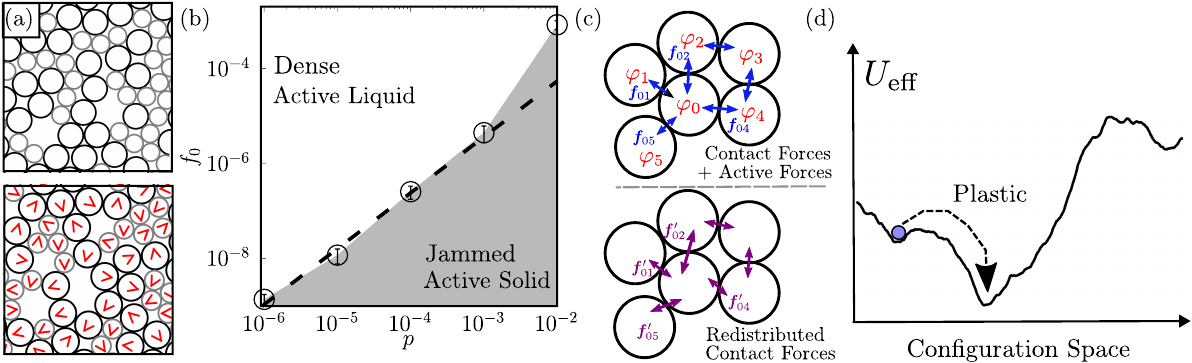}
    \caption{\textit{Yielding, force-balance and stability in infinitely persistent active matter.} (a) \textit{Top:} A jammed packing of soft bidisperse particles. \textit{Bottom:} Under the effect of spatially uncorrelated active forces (red arrows), the contact forces get into mechanical equilibrium with the imposed active forces. (b) In the phase space spanned by pressure ($p$) and active forces ($f_0$), the circles are the critical active forces, $f_c$, that mark the boundary between the active solid and active liquid. At small pressures, $f_c$ scales as $f_c\sim p^{1.17}$ (dashed-line) for system size $N=8192$. See Sec.~\ref{unjam}. Error bars are scaled up by a factor of five for visibility. c) \textit{Top:} The contact forces $\bm{f}$ (blue-arrows) themselves are not in force-balance in active solids. \textit{Bottom:} Using an auxiliary field $\phi$ defined on the particles, we redistribute the contact and active forces to construct a unique network of modified contact forces $\bm{f'}$ (purple-arrows) that are in force-balance. We detail this method in Sec.~\ref{refo}. d) In the limit of infinite persistence time, an effective potential energy $U_\text{eff}$ governs the deformation dynamics in active solids. See Sec.~\ref{macro}.
    } \label{1panel}
\end{figure*}

\section{Model and Methods}\label{methods}
In $d=2$ dimensions, contained within a box of length $L$, we simulate $N$ number of soft discs. See Fig.~\ref{1panel}(a). We use a 50:50 bidisperse system with disc diameter ratio 1:1.4 to prevent crystallization~\cite{o2003jamming,xu2005random}. The steric repulsions between the particles is implemented by a truncated harmonic contact potential, 
\begin{equation}\label{potential}
    U(r_{ij})=\frac{\epsilon}{2}\left(1-\frac{r_{ij}}{\sigma_{ij}}\right)^2\;\Theta(\sigma_{ij}-r_{ij})
\end{equation}
where $r_{ij}=|\bm{r}_i-\bm{r}_j|$ is the interparticle distance, $\bm{r}_i$, $\bm{r}_j$ are the position vectors for the $i^\text{th}$, $j^\text{th}$ particle respectively,  $\sigma_{ij}=R_i+R_j$ with $R_i$ being the radius of the $i^\text{th}$ disc and the Heaviside step function $\Theta(x)=1$ for $x\geq0$ and zero otherwise. 

The dynamics of the particles under the influence of these pairwise steric interactions and active forces is governed by the overdamped equations of motion,
\begin{equation}
\dot{\bm{r}}_i = \frac{1}{\gamma}\left[ -\sum_{j=1}^{N}{\bf \nabla}_{i}U({r}_{ij})+f_0\,{\bf{\hat n}}_{i} \right], 
\label{browneq}
\end{equation}
where $f_0$ is the magnitude of the self-propulsion force and ${\bf{\hat n}}$ is a unit vector which sets the direction of the propulsion force.  The angles these directions make with the horizontal axis are chosen from a uniform probability distribution between $[-\pi,\pi]$. See Fig.~\ref{1panel}(a). These directions remain fixed as we take the rotational persistence time to be $\tau_p=\infty$. Thus, there is no rotational dynamics in this system. We ensure $\sum_{i=1}^N\bf{\hat{n}}_i=\bf{0}$ to avoid displacements of the center of mass of the system~\cite{morse2021}. For all our simulations we implement periodic boundary conditions in both directions. 

In the context of a force-based picture, an active jammed state is one in which the net elastic (steric) force on each particle exactly balances the active force applied to it. Equivalently, in the energy picture, jamming corresponds to the minimization of an \textit{effective} potential energy. This is akin to minimization of enthalpy in the context of jammed systems under compression $H=U+pV$~\cite{xu2017} or under shear stress $H=U-\sigma\gamma V$~\cite{liu2014finite}. This allows us to define the effective potential energy as,
\begin{equation}\label{pe}
U_{\text{eff}}(\{\bm{r}\})=U(r_{ij}) - f_0\,\sum_{i=1}^N\bf{\hat{n}}_i\cdot\bf{r}_i,
\end{equation}
which an infinitely persistent active system is expected to minimize in a jammed state~\cite{liao2018criticality}. Note that such a description is possible only because the particles lack  rotational dynamics which ensures that the self-propulsion directions remain constant in time. The choice of $\sum_{i=1}^N\bf{\hat{n}}_i=\bf{0}$ makes $U_\text{eff}$ translationally invariant {\it{i.e.}} $U_\text{eff}(\bm{r}+\bm{\Delta r})=U_\text{eff}(\bm{r})$ by virtue of the contact potential energy itself being translationally invariant and vanishing of the sum $f_0\bm{\Delta r}\cdot\sum_{i=1}^N\bf{\hat{n}}_i$. For calculating $U_\text{eff}$, we take the position coordinates $\bm{r}_i$ to be the unwrapped coordinates while periodic boundary conditions are used for calculating the interparticle distance $r_{ij}$.

To minimize $U_\text{eff}$, we use the FIRE algorithm~\cite{bitzek2006structural} or Brownian dynamics. The system sizes employed in this study are $N=1024,4096,8192$. We report our results in units of $m$, $\sigma$ (the diameter of the smaller particle), $\sqrt{m\sigma^2/\epsilon}$ for mass, length and time respectively.
The translational friction coefficient $\gamma$ is set to unity. 
Our initial configurations are an ensemble of around $100$ configurations that are jammed at $f_0=0$ and have a fixed pressure $p$. For a given pressure, these configurations will have a distribution of packing fractions $\phi\equiv N\pi (1+1.4^2)/(8 L^2)$. Prior to adding activity, we remove all passive rattlers until we have a single percolating cluster of forces. 

\section{The unjamming transition}\label{unjam}
\subsection{Active danglers}
In passive systems, rattlers are defined to be those particles with coordination number $z<d+1$ where $d$ is the embedding dimension. In the presence of activity, a rattler becomes dynamic and can remain static only when it hits other particles which offers some resistance via the elastic contact forces. However, activity enables the creation of a novel species of particles with $z=2$ which we term as \textit{active danglers}. See Fig.~\ref{2panel}(a). These danglers get stuck in the crevice of two particles where the elastic interaction forces are balanced by the active force of the dangler. 

\subsection{Criterion for yielding}\label{crit}
For finite rotational persistence times, the system can only jam intermittently. To study truly jammed systems, we need systems at zero temperature and at the persistence time of the rotational diffusion $\tau_p=\infty$~\cite{liao2018criticality,mandal2020extreme}. 
In such a way, once the dense packing of active particles are jammed, they remain so, indefinitely, preserving the force-balance.  

In athermal active systems, at $\tau_p=\infty$, the yielding point has been identified to be the state where the applied active force is just enough for the system to become isostatic {\it{i.e.}} $\langle\Delta z\rangle=0$~\cite{anand2024active} where $\Delta z\equiv z-2\,d$. Such a measure calculates the critical force to scale with pressure as $f_c\sim p^{1.5}$. In ref.~\cite{mandal2020extreme}, a different criterion was used (based on kinetic energy and variance of forces) to identify a similar yielding (unjamming) point. In athermal systems, yielding (melting) has been defined as the active force needed for the average displacement of the particles to be of the order of the system size~\cite{liao2018criticality}. Here, the critical active force $f_c\sim \Delta \phi$ and  since $\Delta \phi\sim p$ near the unjamming point~\cite{chaudhuri2010jamming}, we can expect $f_c\sim p$.

Our definition of yielding is relatively stricter. To find the critical force $f_c$, we start with a  passively jammed configuration with no rattlers. Then we add self-propulsion force $f_0$ and minimize $U_\text{eff}$ using FIRE minimization.
If the average speed of the system is less than $v_0=10^{-10}$, we consider the system to be jammed. Since, it is not practical to wait forever for the system to jam, we set a cut-off of $10^6$ time steps of step size $\Delta t=0.2$. If the system gets jammed within this cut-off time, we increase the active force by a small amount $\delta f_0$ and repeat the procedure. See Fig.~\ref{runs} for a representative of the minimization procedure. Past the cut-off time, if the speed of the system is greater than $v_0$, the system is considered to have yielded and the active force used is considered to be the critical force $f_c$ for that particular configuration. The active yielding point $f_c$ is calculated up to a precision of $\delta f_0$ such that $\delta f_c/p=10^{-4}$. This is achieved by first locating $f_c$ to some precision and then iteratively finding $f_c$ at higher precisions around these previously calculated critical force values. We also make sure that we do not have active danglers. When we do have them, we remove them and re-minimize $U_\text{eff}$, until we have configurations in force-balance devoid of active danglers. The change in the packing-fraction distributions is quite minimal. See Fig.~\ref{runs}(b). 

\subsection{The unjamming line}\label{unli}
For small pressures, we see a power-law dependence of the critical force $f_c$ on the pressure $p$ as $f_c(p)\sim p^{\alpha}$. For the largest size ($N=8192$), we find the exponent $\alpha=1.17$ as shown in Fig.~\ref{1panel}\,(b). 
In the thermodynamic limit, we expect $\alpha$ to reach the value of unity as seen in Ref.~\cite{liao2018criticality} (see Fig.~\ref{fc_finite}) for the critical active forces at different system-sizes).
This is based on the expectation that the strength of the critical active force should be proportional to the average of the contact forces in the system. As the contact forces (or the overlap between the particles) are proportional to the (virial) pressure, we expect the critical active force to scale linearly with the pressure. 
Note that the pressure that we refer to here is that of the initial configuration of the passive jammed solid. See Fig.~\ref{active_pressure} where we calculate the `active pressure' using the Irving-Kirkwood expression~\cite{yang2014}.

While $p=0$, describes the passive unjamming point, in the presence of the (infinitely persistent) active forces we have an unjamming line in the ($f_0-p$) plane. The criticality of the unjamming point at $p=0$ is retained as evidenced by the power-law scaling, at least at small pressures. Now that we know the functional dependence of the active yielding force on pressure, we move onto studying the force networks of these forced balanced active  jammed states by analysing the force distributions at $f_0< f_c$. 

\begin{figure*}[t]
    \centering
    \includegraphics[width=0.95\textwidth]{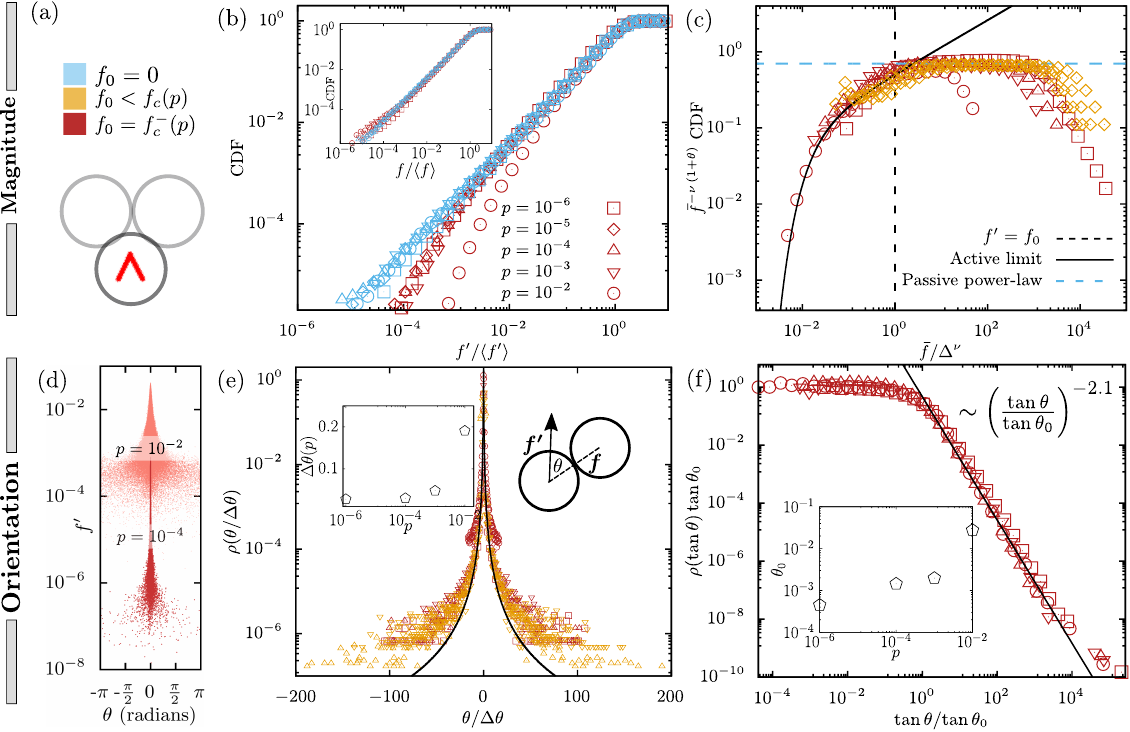}
    \caption{\textit{Universal scaling of force distributions in jammed active solids.}
    (a) active danglers ($z=2$) such as the black particle whose active force is balanced by elastic forces of the two gray particles are systematically removed prior to force analysis.  (b) Inset: The contact forces $f$ for the passive jammed systems (blue) obey power-law distributions at $f/\langle f\rangle<1$ unlike the active systems with $f_0\rightarrow f_c^-$ (red) at small $f$. On redistributing the contact and active forces into a network of  redistributed forces $f^{\prime}$, a decent collapse is obtained at small pressures $p$. (c) The scaling law (Eq.~\ref{ansatz}) is applicable in the entirety of jammed active solid phase (Fig.~\ref{1panel}) - at all pressures $p$ and active forces including $f_0<f_c$ (orange). 
    (d) The scatter plot of $f^{\prime}$ and the angle $\theta$ it makes with the bond vector. (e) The normalized density distributions of $\theta$ has a cusp-like shape at $\theta\rightarrow0$ described by $\rho(\theta/\Delta \theta)\sim (\theta/\Delta \theta + c)^{-2.9}$ with $c=0.23$ (black line). The inset shows the standard deviation $\Delta \theta$ monotonically increasing with $p$. (f) The collapse of density distributions of $\text{tan}\,\theta=f_\perp/f_\parallel$, the ratio of the tangential and the normal components of $f^\prime$ has a plateau for $\text{tan}\,\theta/\text{tan}\,\theta_0\ll1$ and a power-law at large values. The inset is the normalization angle $\theta_0(p)$ monotonically increasing with $p$ akin to $\Delta \theta(p)$ in (e). System size, $N=8192\,(4096)$ for top (bottom) respectively.
    } \label{2panel} 
\end{figure*}

\section{Microscopic force distributions}\label{forceDist}
\subsection{Bare contact forces}
An amorphous solid responds to stress in a highly heterogeneous manner by forming force chains~\cite{liu1995force,coppersmith1996model}. These force chains which construct the force network in the system are central to our understanding of the mechanical properties and memory effects of passive jammed systems~\cite{tighe2010}. Near the unjamming point, the distributions of the magnitude of the interaction forces are not a simple Gaussian-like distribution peaked at its average. Rather, 
for contact forces much larger than the average contact force ($f\gg \langle f\rangle$), the distribution decays exponentially~\cite{o2001force}. For $f\ll \langle f\rangle$, they exhibit a power-law distribution
of the form $\rho\,(f)\sim f^{\theta_l}$ ~\cite{lerner2012toward} with $\theta_l$ being the corresponding power-law exponent.  In Sec.~\ref{unjam}, we extended the passive unjamming point at $p=0$ to the unjamming line in the $(f_0-p)$ plane. Now, we ask, is there a universal scaling form for the force distributions in jammed active systems near the unjamming line? 

The power-law distribution implies an existence of a large number of contacts with very weak contact forces, as is shown via the blue data points in the inset of Fig.~\ref{2panel}\,(b) which displays the cumulative distributive function (CDF) of the contact forces $f$ normalized by their averages. It is natural to expect that the addition of active forces could potentially modify these weak contacts. The question is whether the force distribution can still be described by a similar scaling law or will it change completely. 
We observe that for small contact forces in jammed active systems, we lose the collapse in the force-distributions. See the red data
points of the inset of Fig. ~\ref{2panel}\,(b).

However, this does not represent the complete picture of the microscopic forces in jammed active systems. While in the passive case, contact forces by themselves are in exact force-balance, that is not the case for our persistent active solid. We need to look at the force-balance formed by the bare contact forces and active forces. Constructing an equivalent force-balanced network that incorporates both active and elastic contributions is non-trivial. We address these questions in the following section.

\subsection{Redistributed forces}\label{refo}
To analyse the statistics of all the microscopic forces in the system, we employ a Laplacian framework which introduces local force-balance on particles by allowing for the presence of non-contact body forces~\cite{ramola2017stress}. As the active forces on the particles here are body-forces, we use this framework to redistribute the active forces and generate a modified force distribution $f^{\prime}_{ij}$ between the particles in contact. This new force network accounts for both the repulsive harmonic interaction $f_{ij}$ and active forces $f_0\,\bm{\hat{n}}_i$ on particles. We briefly describe this scheme below.

We use bold lower case and bold upper case to denote 2-dimensional vectors and $N\times N$ dimensional matrices respectively.
To represent $N$ dimensional vectors whose entries themselves are 2-dimensional vectors, we use the Dirac notation (denoted with $|\bm{..}\rangle$). For a network of contacts, the graph Laplacian, a $N\times N$ matrix is defined as $\bm{\Box}^2 \equiv \bm{A-D}$
where $\bm{A}$ is the adjacency matrix of the contact network and $\bm{D}$ is a diagonal matrix which encodes the number of contacts on each particle~\cite{newman2018}. We define a 2 dimensional auxiliary field vector $\bm{\varphi}$ on each particle (see the bottom part of Fig.~\ref{1panel}(c)) such that it satisfies the relation $\bm{\Box}^2 |\bm{\varphi}\rangle=-f_0\, |\bm{n}\rangle$ where the right-hand side is the vector that contains the body forces on each particle. By inverting the graph Laplacian, we have $\varphi=-f_0 \,\bm{\Box}^{-2}|\bm{n}\rangle$. The  redistributed force at every contact 
\begin{equation}\label{redistributed_f}
 |\bm{f}^{\prime}\rangle=|\bm{f}\rangle-\bm{\Box}\, |\bm{\varphi}\rangle.
\end{equation}
is the sum of the original contact force and  the gradient $\bm{\varphi}$ of the particles sharing the contact. Thus, the two sets of forces (original contact forces and active forces) are unified into a new set of  contact forces. This redistributed force network can be interpreted as the active analogue of the bare contact network in a passive system. 
The configuration which is held in mechanical equilibrium by the combination of active forces and contact forces, will \textit{also} be held in equilibrium by the redistributed contact force. 

Note that this set of  redistributed contact forces (Eq.~\ref{redistributed_f}) satisfy Newton's third law since both $|\bm{f}\rangle$ and $\bm{\Box}\,|\bm{\phi}\rangle$ change sign on swapping the particle indices .  The  redistributed contact forces are unique for a given contact force network and body forces. Their magnitude in general is different from the contact forces and they are not constrained to be a radial force ({\it{e.g.}} they can have transverse components). More details can be found in~\cite{ramola2017stress}.

\subsubsection{Magnitude}
We now calculate the statistics of the magnitude of the redistributed forces. We observe that the presence of the active danglers gives rise to a plateau at $f'\sim f_0$ in the cumulative force distributions; see Fig.~\ref{plateau}(a). We also notice that this plateau becomes insignificant at large system sizes; see Fig.~\ref{plateau}(b). To get rid of this  finite-size artifact, we remove the active danglers in an iterative fashion as explained in Sec.~\ref{crit}. All data in this section is generated following this protocol.

We first investigate the redistributed forces of configurations near the unjamming line, $f_0=f_c(p)$ of the phase diagram in Fig.~\ref{1panel}(b).  We observe that for small pressures, the cumulative force distributions collapse fairly well for $p <10^{-2}$. See Fig.~\ref{2panel}(b). It is quite interesting that the construction of modified contact forces, which accounts for all the bare contact and active forces in the system, ensures a collapse of the contact force distributions. However, at a large pressure of $p=10^{-2}$, the distribution peels away from the rest of the collapsed data. 
For the passive force distributions in Fig.~\ref{2panel}(b), the redistributed force $f'$ simply reduces to the contact force $f$ itself.

To capture the effect of activity in the bulk of the jammed active phase (Fig.~\ref{1panel}(b)), {\it{i.e.}} for all pressures and  all magnitudes of activity that retain the jammed phase $f_0<f_c(p)$, we propose a scaling ansatz of the form,
\begin{equation}\label{ansatz}
\text{CDF}(\bar{f})\sim\bar{f}^{\,\nu(1+\theta_l)}\;g\,(\bar{f}/\Delta^{\nu}), 
\end{equation}
with $\bar{f}\equiv f^{\prime}/\left<f^{\prime}\right>$ and $\Delta\equiv f_0/\left<f^{\prime}\right>$. 
We see an excellent collapse of the force distribution curve with exponent $\nu=1.0$. See the red data points in Fig.~\ref{2panel}(c). Thus, the reduced variable $y=\bar{f}/\Delta^{\nu}$ is essentially $f^{\prime}/f_0$.
Furthermore, the scaling ansatz holds true for the force distributions even below the unjamming line ($f_0<f_c(p)$). See the orange data points in the same figure, where we have the force distributions at $f_0/f_c=0.2,0.4,0.6,0.8$ for the configurations with initial pressure $p=10^{-4}$. 

The nature of the collapse is qualitatively different across $f^\prime=f_0$, represented by a vertical dashed line in Fig.~\ref{2panel}(c). 
In the limit of $y \to \infty$, the scaling function $g(y)$ reduces to $\,g(y)\sim c$ where $c$ is a constant, recovering the passive scaling relation $\text{CDF}(\bar{f})\sim \bar{f}^{\,1+\theta_l}$ with the exponent $\theta_l=0.15$~\cite{lerner2013low}. The passive regime ($f^\prime>f_0$) is indicated by the horizontal blue dashed-line in the figure. 
In the other limit, $y\rightarrow0$, $g(y)\sim \text{exp}(-y/y_c)\,y^{\alpha-(1+\theta_l)}$, which is the solid black line fitting the active regime ($f^\prime<f_0$). The fit parameters $\alpha=0.35$ and $y_c=0.017$. From the data we have, we cannot conclude if this is the unique scaling form. A scaling form with a gapped distribution in the active regime instead of the exponential decay fits this data equally well. See sec.~\ref{sirf}. We do not see any strong finite-size dependence within the sizes we studied; see Fig.~\ref{finite_size}(a). 

The process of redistribution of contact and active forces allows us to see a qualitative difference in the distribution of contact forces for $f^\prime<f_0$. Moreover, the statistics of the magnitude of forces display a universality in the force distributions which holds in the entirety of the bulk of the jammed active solid regime {\it{i.e.}} for all pressures and active forces within the grey-shaded area indicated in Fig.~\ref{1panel}(b).  

\subsubsection{Orientation}
The redistributed contact forces $\bm{f}^\prime$ are not constrained to be along the bond-vector between particles in contact unlike the contact forces $\bm{f}$. The forces $\bm{f}^\prime$ and $\bm{f}$ cease to be parallel due to the gradient of the auxiliary field $\bm{\phi}$ which in general will not be along the bond vectors (see Eq. \ref{redistributed_f}).  
The pairwise redistributed forces for the jammed active solids, can have not just radial components but also transverse components. We now analyse the bond-angle $\theta$ these forces make with the bond vector between particles in contact. 
See Fig.~\ref{2panel}\,(d) for a scatter plot of $\theta$ and $f^{\prime}$. For larger forces at a given pressure, we see that $\theta\approx0$ as is expected for passive systems. For smaller forces, there is a wide distribution of angles. The probability distributions for the normalized angles where $\theta$ is normalized by $\Delta \theta$ (the standard deviation of the distribution) displays a cusp-like shape as $\theta\rightarrow0$. See Fig.~\ref{2panel}(e).  For the system size of $N=4096$ at different pressures calculated at $f_0=f_c$, the probability distributions have a decent collapse for small pressures. This indicates that while the ensemble average $\langle\theta\rangle$ is zero, the only relevant parameter that determines the distribution is the standard deviation $\Delta\theta(p)$ which is a monotonically increasing function of pressure $p$; see inset of Fig.~\ref{2panel}(e). 

\begin{figure*}
    \centering
    \includegraphics[width=0.9\textwidth]{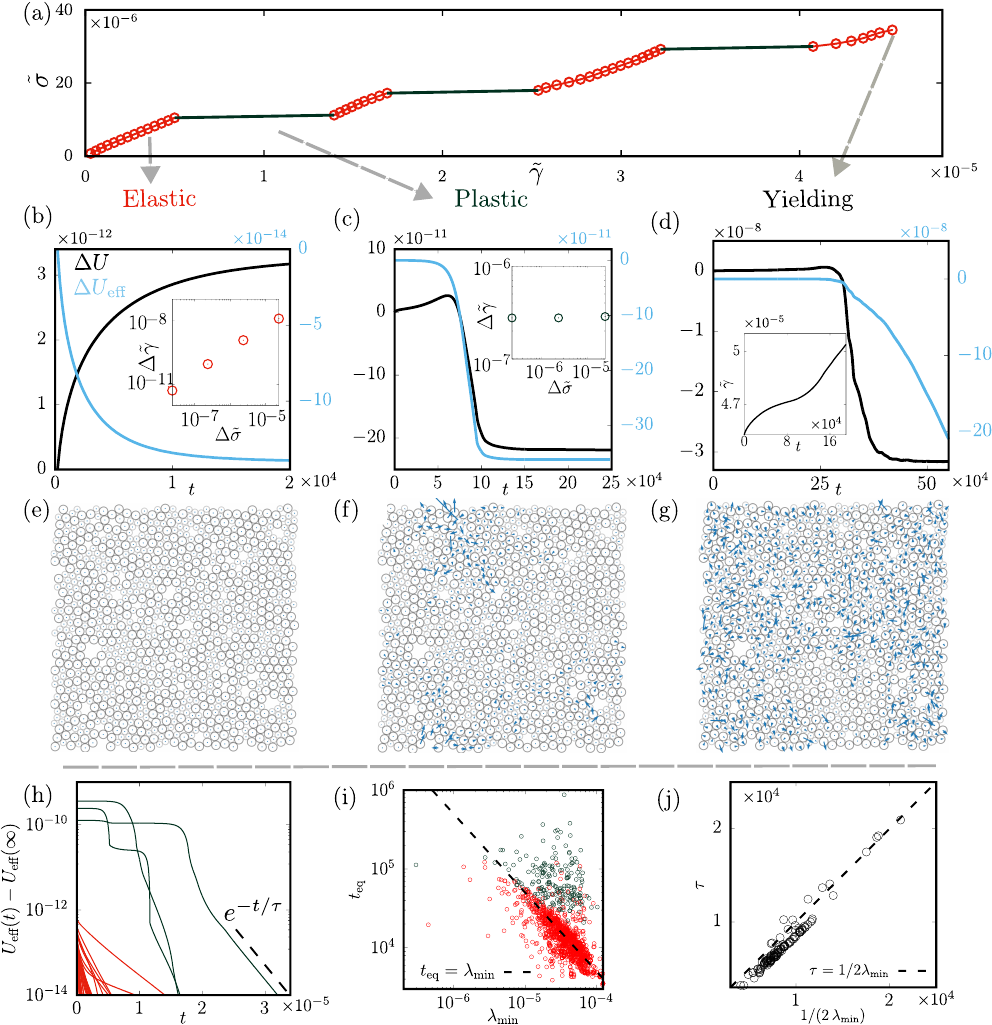}
    \caption{\textit{Elasticity, plasticity and yielding in jammed active solids.} (a) The typical strain response during a quasistatic increase in imposed active force, using $\delta{f_0}=10^{-7}$, on a configuration of jammed particles displays elastic deformations (red) punctuated by plastic events (dark-green) and terminates by yielding.  (b,c,d) Typical time evolution of the contact potential energy $\Delta U(r_{ij})$ (black) and effective potential energy $\Delta U_\text{eff}$ (blue) during elastic, plastic periods and yielding. (Insets of b,c,d) The change in active strain $\Delta \tilde {\gamma}$ scales with the change in active stress $\Delta \tilde {\sigma}$ for elastic events in contrast to plastic events. At yielding, strain increases monotonically with time.  
    (e,f,g) Displacement fields (vectors scaled up by a factor of $3\times10^2$ for visibility) during a typical elastic, plastic and yielding event. 
    (h) The relaxation dynamics of the effective potential energy for elastic (red) and plastic events (dark green) have exponential form at long times.
    (i) The inverse relationship between equilibration time $t_\text{eq}$ and the smallest non-trivial eigenvalue of the Hessian $\lambda_\text{min}$ can be seen for the elastic events (red). The plastic events lie above the line of $t_\text{eq}=\lambda_\text{min}^{-1}$ as the time to relax is typically longer in plastic events as seen in the dark green curves of (h). 
    (j) The numerically calculated time constant at long times, $\tau$, match with $1/(2\lambda_\text{min})$. All figures use $N=1024$ and { $p=10^{-4}$}.}\label{3panel}
\end{figure*}

Transverse forces are also known to occur in the context of jammed frictional passive systems where the force-balance is between the normal elastic and tangential frictional forces. In jammed active systems, the redistributed force between two particles in a contact can be written as $\bm{f^{\prime}}=\left(f_{\parallel},f_{\perp}\right)$ where the parallel and perpendicular components are with respect to the bond vector $\bm{r}_{ij}$. We consider the ratio of these two forces $f_\perp/f_\parallel=(f^{\prime}\, \text{sin}\,{\theta})/(f^{\prime} \,\text{cos}\,{\theta})=\text{tan}\,{\theta}$. See Fig.~\ref{2panel}\,(f), where we show the distributions of $\text{tan}\,\theta$ normalized by $\text{tan}\,\theta_0$ such that $\rho(\text{tan}\,\theta)\,\text{tan}\,\theta_0$ is unity as $\text{tan}\,\theta/\text{tan}\,\theta_0\rightarrow0$. We observe a plateau in the distributions for $\text{tan}\,\theta<\text{tan}\,\theta_0$ i.e. when the bond-angles are almost passive-like ($\theta=0$). However, for large bond-angles, we see a clear power law behaviour: $\rho(\text{tan}\,\theta)\,\text{tan}\,\theta_0 \sim (\text{tan}\,\theta/\text{tan}\,\theta_0)^{-2.1}$. 

This is true even for $f_0<f_c(p)$ (see Fig.~\ref{finite_size})(b). This is interesting since in passive frictional systems, the probability distribution of normalized tangential forces decays exponentially~\cite{silbert2002statistics,majmudar2005contact}. This suggests that the (redistributed) force distributions of jammed active systems are unique and cannot be trivially mapped to an effective friction in passive jammed solids. 

\section{Macroscopic response to stress}\label{macro}
We will now look at the macroscopic stress-strain curves which reveal different types of response to stress in the solid phase~\cite{morse2021,berthier2025yielding}. 
We will also look at the Hessian and its predictive agency in determining the dynamics of the system as it approaches a mechanical equilibrium.

In this section, we impose a quasistatic ramping of active force, using $\delta{f_0}=10^{-7}$, with the effective energy $U_\text{eff}$ being minimized at every ramping step. We use athermal Brownian dynamics (Eq.~\ref{browneq}) instead of FIRE~\cite{bitzek2006structural} to avoid jumps out of the local minimum caused by inertia. Note that we can reach jammed states with considerably larger active forces for Brownian dynamics in comparison with FIRE (see Fig.~\ref{fc_finite}). In this exercise, we do not remove active danglers while minimizing,  since we would like to compare Hessians at different activities by retaining the original dimensionality of the Hessian.

\subsection{Elasticity, plasticity and yielding}
Increasing the active force $f_0$ amounts to applying local stresses on the particles along the directions ${\bf{\hat{n}}}_i$. The particles respond to this stress by undergoing a finite amount of displacement (strain). In a parallel approach, one can displace the particles along the directions ${\bf{\hat{n}}}_i$ by a finite amount and in response, the system generates local stresses. Controlling the strain or stress are referred to as athermal quasistatic random displacements (AQRD) or athermal quasistatic random forces (AQRF) respectively~\cite{morse2021}. For AQRF, stress and strain can be defined as,
\begin{equation}
\begin{split}
    \tilde{\sigma}&= \frac{f_0}{L} \frac{N}{\sqrt{12}}\\
    \tilde{\gamma}&= \frac{\langle \bm{n} | \bm{\Delta x}\rangle}{L} \frac{\sqrt{12}}{N}.
    \label{siggama}
\end{split}
\end{equation}
where $|\bm{\Delta x}\rangle$ is the particle displacement vector of the system. 
In Fig.~\ref{3panel}(a), we see that on continuous quasistatic increase of imposed active stress, the system responds either by elastically deforming, suffering a plastic rearrangement of particles and eventually yields. The response seen here is very similar to the results obtained in Ref.~\cite{morse2021}. Few representative displacement fields are shown in Fig.~\ref{3panel}(e-g).

Elastic deformations are reversible in nature and the strain increases proportionally with imposed active stress. On the other hand, the irreversible plastic deformations are accompanied by large jumps in the strain. 
The change in strain $\Delta \tilde{\gamma}$ in elastic deformations scales with change in imposed stress $\Delta \tilde{\sigma}$; see inset of Fig.~\ref{3panel}(b). However, across the plastic event, the change in strain is independent of $\Delta \tilde{\sigma}$; see inset of Fig.~\ref{3panel}(c).
Moreover, the plastic events are necessarily accompanied by a change in the contact network of the particles (Ref.~\cite{xu2023discontinuous} discusses the case for harmonic interactions). 
We could verify that the discontinuities in the stress-strain curve are also accompanied by an abrupt change in the values of equilibrated contact potential energy $U$, effective potential energy $U_\text{eff}$ and the smallest eigenvalue (discussed in more detail later on). Thus, this paints the picture of the system departing from one minimum of $U_\text{eff}$ to find another in response to a small increase in the imposed active stress, as long as $f_0 < f_c(p)$. 

The system yields for $f_0>f_c(p)$, when the system cannot find a new minimum of $U_\text{eff}$ within our observation time window. In the steady state of the fluid phase, spatial correlations of both force directions and velocities $d\bm{r}_i/dt$ become long-ranged~\cite{henkes2020}.
To explain the change in $U_\text{eff}$ in the fluid phase, we ignore such correlations, and take the velocity to be a scalar multiple of the active director vector $d\bm{r}_i/dt = v_0 \, \bm{\hat{n}}_i$ where $v_0=f_0/\gamma_{\rm eff}$ with $\gamma_{\rm eff}$ being the effective drag coefficient of the active fluid. The effective potential energy then changes with time as, $U_\text{eff}(t)\approx - f_0^2 t/\gamma_{\rm eff}$ \textit{i.e.} it linearly decreases with time. This is demonstrated in the late-time dynamics shown in Fig.~\ref{3panel}(d). The strain is also seen to linearly increase with time as can be seen in the inset of this figure, i.e. a steady flow with a finite strain rate emerges~\cite{villarroel2021critical}.

\subsection{The Hessian and relaxation dynamics}
In athermal jammed systems, relaxation following a perturbation is governed by the local curvatures of the potential energy landscape. Here, we infinitesimally perturb a jammed state of athermal active particles and investigate how the resulting relaxation dynamics are controlled by the Hessian of the effective potential energy $U_\text{eff}$. 
The Hessian for our jammed active system is defined as, 
\begin{equation}
    \bm{K}_{ij}=\frac{\partial^2 U_{\text{eff}}}{\partial \bm{r}_i \partial\bm{r}_j}\bigg|_{\bm{r}=\bm{r}_\text{eq}}=\frac{\partial^2 U}{\partial \bm{r}_i \partial\bm{r}_j}\bigg|_{\bm{r}=\bm{r}_\text{eq}}.
\end{equation}
The double derivative of the $U_\text{eff}$ reduces to the double derivative of $U$ due to the active term being only linearly dependent on positions. 
Activity, therefore, can only affect the Hessian implicitly since it changes the equilibrium position ($\bm{r}_\text{eq}$) of the particles, thereby changing the point in the configuration space at which the Hessian is evaluated. 

In passive jammed systems, with decreasing packing fraction (above $\phi>\phi_J$), the smallest nontrivial eigenvalue of the Hessian, $\lambda_\text{min}$ continuously goes to zero as the system approaches the unjamming transition~\cite{ikeda2020}. The same holds true for athermal particulate systems under simple shear approaching a pstrain~\cite{maloney2006,manning2011} or stress~\cite{xu2023discontinuous} unlike the  frictional jammed systems~\cite{ishima2023} where one observes discrete and abrupt jumps in the eigenvalues and the instability cannot be predicted.  

We find that, at the plastic instability, the increment in active forcing leads to sudden changes in the eigenvalues of the Hessian (see Fig.~\ref{ev_fine}). The smallest eigenvalue does not continuously go to zero when the system approaches the instability. Hence the location of the plastic instability cannot be predicted from the eigenvalues of the Hessian.  This is a consequence of the choice of using harmonic interactions which necessarily implies that the curvature of the potential does not vanish at the point of the contact. The system suffers  discontinuous jumps in strain ($\tilde{\gamma}$) where $ \lambda_\text{min}$ jumps across this instability rather than continuously approaching zero at that transition~\cite{xu2023discontinuous}. Thus, the Hessian can not be used to predict an impending plastic event. However, by using Hertzian interactions instead, we do observe a continuous decrease of $\lambda_\text{min}$ to zero (data not shown) at the jumps.

In the elastic branch, when a small activity is imparted on each particle, the system relaxes to a new minimum (close to the previous minimum) of $U_\text{eff}$. In the approach to mechanical equilibrium, the potential energies $U(r_{ij})$ and $U_\text{eff}$ monotonically increase and decrease respectively, as can be seen in  Fig.~\ref{3panel}\,(b). 
The relaxation dynamics of the system is determined by the Hessian. The vibrational mode, corresponding to the $k^\text{th}$ largest eigenvalue of the Hessian, is expected to relax as $\sim \text{exp}\,(-2\lambda_k t)$.
At long times, when all other modes have relaxed, the equilibration dynamics is determined by the vibrational mode with the smallest eigenvalue $\lambda_\text{min}$. In this long time limit, $\Delta U_\text{eff}(t)\equiv U_\text{eff}(t)-U_\text{eff}(\infty) \sim \text{exp}\,(-2\lambda_\text{min}t)$~\cite{ikeda2020}. Thus, the relaxation constant in the long time limit is $1/(2\lambda_\text{min})$ implying a much larger equilibration time for systems which are approaching a minimum with smaller $\lambda_\text{min}$. In our jammed active systems too, we observe the exponential nature of the relaxation at long times both for elastic and plastic events (shown in red and dark-green curves respectively in Fig.~\ref{3panel}(h)).

Given that our $\lambda_\text{min}$ does not continuously go to zero (due to the harmonic interactions~\cite{xu2023discontinuous}) we do not expect a divergence of the timescale associated with the exponential relaxation, as the system approaches a plastic event. However, the inverse relation between the relaxation time $t_\text{eq}$ and the smallest nontrivial eigenvalue of the Hessian, $\lambda_\text{min}$ still holds (see Fig.~\ref{3panel}(i)). {Here, $t_\text{eq}$ is the total time it takes for the average speed of the particles to decrease to less than $v_0=10^{-10}$ as mentioned in Sec.~\ref{methods}.} The inverse relation does not hold well for the plastic events when the system moves from one `potential well' to another (dark green dots have displacements $\Delta x$ larger than $2\times10^{-4}$) . The time it takes for doing so will be larger than that predicted by the relaxation of the smallest eigenvalue, evident from the relaxation curves corresponding to plastic events in Fig.~\ref{3panel}\,(b) (also in dark-green).
Given that the total time for relaxation includes both the exploration of the landscape and the relaxation of all the vibrational modes, the inverse relationship of $t_\text{eq}$ and $\lambda_\text{min}$ (corresponding to the final minimum approached by the system) is only approximate. 
We find that the relaxation constant of the smallest non-trivial vibrational mode $\tau$, which we measure by fitting the long time dynamics (that describes the relaxation in the final minimum) of $\Delta U_\text{eff}$ with $\text{exp}(-t/\tau)$ (see Fig.~\ref{3panel}(h)), matches well with $1/(2\lambda_\text{min})$ (see Fig.~\ref{3panel}(j)). This holds true both for elastic or plastic events since the relaxation dynamics in the long time limit is determined only by the Hessian of the potential well the system is finally approaching.

\section{Discussion}
Jamming in zero temperature passive systems is a geometrical problem. When the number of geometric constraints are greater than the number of degrees of freedom, the system must be jammed.  The infinite persistent limit lets us write down an effective potential which relaxes as the system responds to an imposed active stress. In this paper, we have carefully analysed  jammed active states of infinitely persistence active matter by studying the distributions of the microscopic contact forces, macroscopic stress-strain response and the relaxations of the system in light of the Hessian of the effective potential energy.

With active forces, force-balance is not satisfied just by the bare contact forces. This poses a fundamental problem in studying the statistics of contact forces since they do not reveal the whole picture of the force-balance in the system. By using the Laplacian framework~\cite{ramola2017stress}, we construct a network of modified forces which by themselves achieve force-balance at the level of every particle by accounting for both the elastic and active forces. 
This redistribution reveals the scaling form obeyed by  the magnitudes of modified force distributions that hold throughout the bulk of the jammed active phase. Moreover, the distribution of the orientation of the redistributed active forces too show a power law distribution. These findings suggest a universal description of force statistics of jammed active solids. 
In general, the modified forces include transverse faces akin to the frictional system of particles.
It will be interesting to explore a quantitative map between active systems and passive frictional systems which too have transverse contact forces.
In force-space of passive jammed systems, the force-balance of contact force on each particle implies that the entire force-space of the system can be tiled by the contact forces by laying down the forces on a particle in a cyclic fashion~\cite{sarkar2013origin}. In the same vein, the force-balance of redistributed forces in jammed active systems implies that these force vectors tile the force space. Geometrical convexity is important in understanding rigidity in a wide-variety of system classes, be it in the force-networks of jammed systems~\cite{sarkar2013origin} or strain-induced transitions in spring networks~\cite{gandikota2022rigidity}. In the jammed active systems we have looked at, the force-tiles  corresponding to redistributed contact forces remain convex - reminiscent of the convexity of the bare contact-forces in the passive jammed systems.

At the level of macroscopic response, jammed active solids display elastic deformations punctuated by plastic events and eventual yielding. The plastic events are abrupt and not preceded by a continuous softening of the Hessian eigenvalues. Yet, the inverse of the smallest eigenvalue of the vibrational modes is seen to predict the time required to relax to an energy minimum, quite well. In thermal suspensions of dense active particles with Lennard-Jones interactions, divergences are observed in the time taken by the system to reach the steady state when the system is near its yielding point~\cite{goswami2025}. Using, harmonically interacting spheres, we do not see any diverging timescale in zero temperature systems as the jumps in the smallest vibrational frequencies are discrete and  the eigenvalues associated with the lowest energy mode, do not continuously go to zero~\cite{xu2023discontinuous}. However, preliminary investigations using Hertzian spheres, which have vanishing stiffness at the point of contact, indicate the smallest vibrational frequency continuously goes to zero as the system approaches an instability. The inverse of this frequency  offers a diverging timescale. 
It could be fruitful to compare the divergences in thermal active systems to the ones we get in athermal systems. We will address this question in a future work. 

We observed that activity creates a new class of particles (active danglers) with coordination number $z=2$. 
An jammed active system cannot have any true rattlers $z<d+1$ since the particles are constantly driven. They can come to rest only when they come in contact with other particles which establishes a force-balance between the particle's active force and the contact forces due to elastic interactions. The active danglers with $z=2$ tend to get stuck at the crevice between two other particles. These anomalies are confined to active systems alone with no counterpart in passive jammed systems. We observe that with larger active forces, we have a larger number of danglers. Moreover, these danglers lead to the creation of a plateau in the modified force distributions at $f^\prime=f_0$. Studying the nature and density of these danglers, their creation in the system as a function of activity and their precise effect on force distributions will be an interesting future direction of investigation. 

Many questions remain in the subject of jammed active systems. Such as, how the history of preparation of jammed systems affects the force distributions in infinitely persistent jammed active systems. If we start from a liquid phase with system-scale velocity correlations, and quench the system to a jammed state, will it change the observed scaling of the force distributions? Recently, the elasticity of the MIPS cluster was probed~\cite{martin2025motility}. In the infinite persistence limit, how do the jammed active states elastically respond to externally imposed strains and how does it relate to the elasticity of the MIPS cluster?
Thus, fundamental questions remain that need to be addressed towards building a more complete understanding of jammed active systems and thereon of dense active systems themselves. We hope that this paper stimulates further work in this direction.
\section*{Acknowledgements}
We thank Surajit Chakraborty for providing us with the passive jammed configurations and Kabir Ramola for the code that calculates modified contact forces. We thank Deshpreet Singh Bedi for carrying out the initial part of this project.  We acknowledge useful discussions with Srikanth Sastry. M.C.G. and C.D. acknowledge support of the Department of Atomic Energy, Government of India, under project no. RTI4019. R.M. acknowledges support from the ANRF, India, through PMECRG (project ANRF/ECRG/2024/002036/PMS). B.C. acknowledges support by NSF-DMR-2026834, and NSF CBET-2228681. P.C. thanks IUSSTF-JC-026-2016 for support.

\bibliography{References}
\bibliographystyle{unsrt}

\onecolumngrid
\vspace{20cm}
\section*{Supplementary Information}
\renewcommand\thefigure{SI.\arabic{figure}} 
\setcounter{figure}{0}

\subsection{Minimization of $U_\text{eff}$}
\begin{figure}[h]
    \centering
    \includegraphics[width=0.75\textwidth]{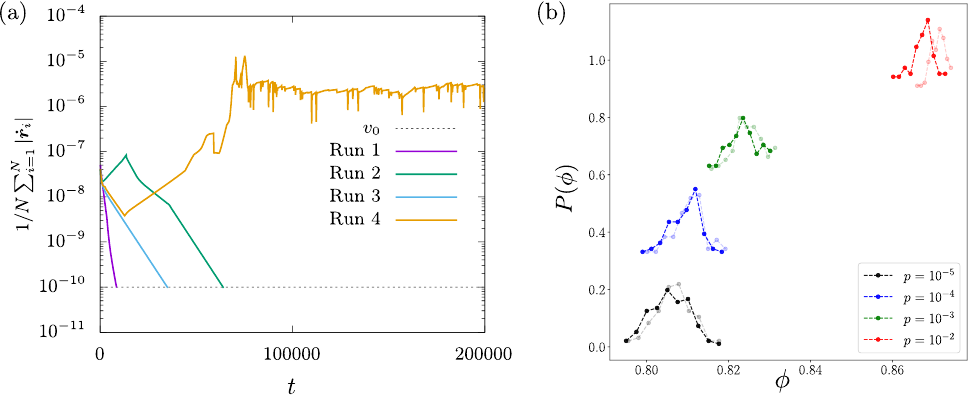}
    \caption{(a) Each run at a fixed active force $f_0$ is a FIRE minimization procedure which terminates when the average speed of the particles is $\le 10^{-10}$. We remove the active danglers with $z=2$ and re-minimize $U_\text{eff}$ at the same $f_0$. (b) The probability distributions of the packing fractions before and after removal of rattlers displayed in translucent and solid line-points respectively for $N=1024$. For visual clarity, $P(\phi)$ of every pressure is offset by a value of 0.3 relative to $P(\phi$) of its immediately lower pressure.}\label{runs}
\end{figure}

\subsection{Finite-size dependence of $f_c$ scaling with pressure $p$}
\begin{figure}[h]
    \centering
    \includegraphics[width=0.45\textwidth]{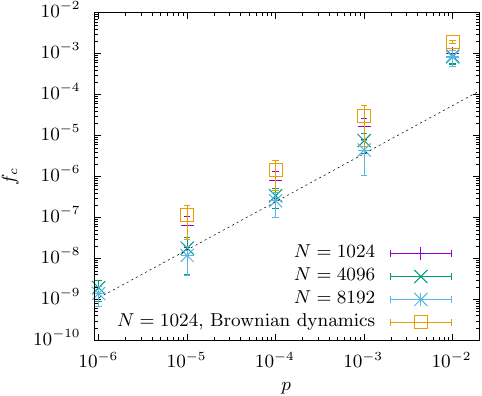}
    \caption{The yielding point $f_c$ at different system sizes. The dotted line is $f_c\sim p^{1.17}$. Note that $f_c$ found using Brownian dynamics is larger than the ones found using the FIRE algorithm.}\label{fc_finite}
\end{figure}

\newpage
\subsection{Active pressure}
\begin{figure}[h]
    \centering
    \includegraphics[width=0.45\textwidth]{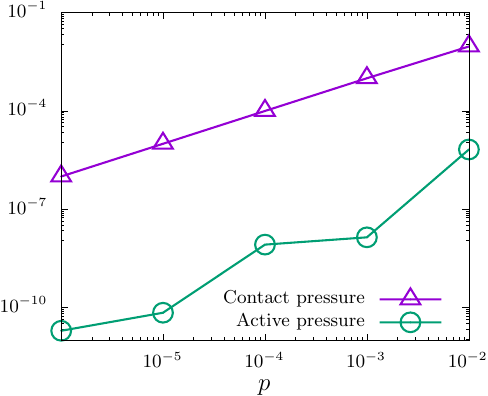}
    \caption{Active pressure at $f_0=f_c^-$ increases with increasing pressure. System size $N=8192$.}\label{active_pressure}
\end{figure}

The stress tensor may be defined using the Irving-Kirkwood expression~\cite{yang2014},
\begin{equation}
    \sigma_{\alpha\beta}^\text{int}=\frac{1}{L^2}\left<\sum_{i\neq j}F_{ij}^{\alpha}\,r_{ij}^{\beta}\right>,     \sigma_{\alpha\beta}^\text{act}=\frac{1}{L^2}\left<\sum_{i}F_{i}^{\alpha}\,r_{i}^{\beta}\right> 
\end{equation}
and the pressure calculated as the trace of these matrices is shown in Fig.~\ref{active_pressure}. As expected, the active pressure is much smaller than the contact pressure. 

\subsection{Redistributed forces}\label{sirf}

\begin{figure}[h]
    \centering
    \includegraphics[width=0.85\textwidth]{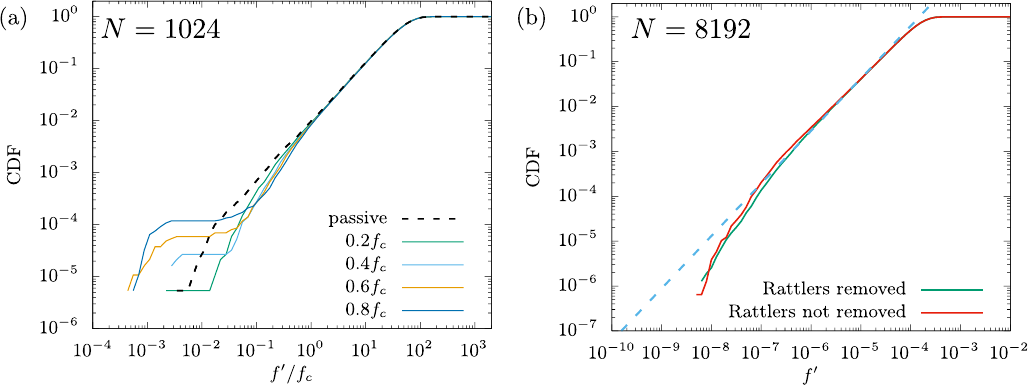}
    \caption{(a) The cumulative distribution functions of redistributed forces $f^\prime$ for pressure $p=10^{-5}$ and systems size $N=1024$ with active force $f_0$ at different fractions of yielding force $f_c$ when we \textit{do not remove} rattlers. (b) The cumulative distribution function of $f^\prime$ near the yielding line ($f_0=f_c(p)$) for $N=8192$ and pressure $p=10^{-4}$.  Note the distinct plateaus at $f^\prime<f_c$ for $N=1024$ which are not present in the larger system with $N=8192$.}\label{plateau}
\end{figure}

\begin{figure*}[h]
    \centering
    \includegraphics[width=0.95\textwidth]{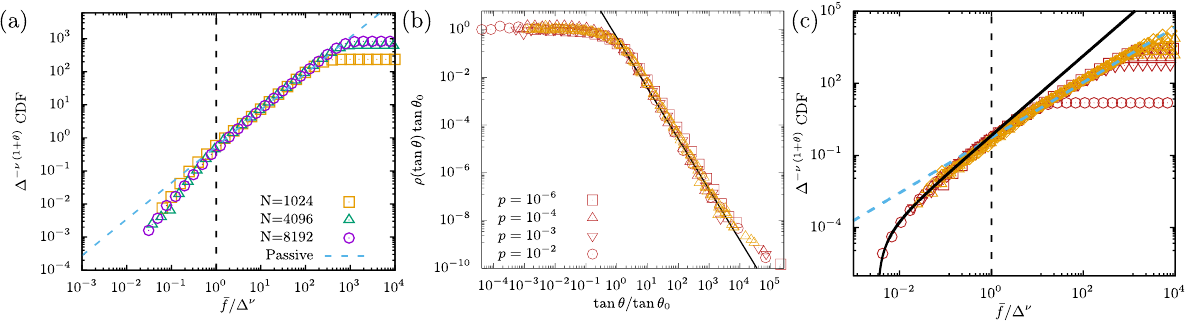}
    \caption{(a) The redistributed contact forces collapsed using the ansatz in Eq.~\ref{ansatz2} at different system sizes. (b) The collapse of density distributions of $\text{tan}\,\theta=f_\perp/f_\parallel$, the ratio of the tangential and the normal components of $f^\prime$ both at $f_0=f_c^-$ (red) at various pressures and at $f_0<f_c$ for $p=10^{-4}$. The collapse is applicable in the entirety of the jammed active solid phase akin to the magnitudes of the redistributed forces. (c) The scaling ansatz (Eq.~\ref{ansatz2}) is applicable in the entirety of jammed active solid phase - at all pressures $p$ and active forces including $f_0<f_c$ (orange). The black solid line defines the functional form of the scaling function in the active regime. Blue dashed line is the passive scaling. System size is $N=4096$ for (b,c).}\label{finite_size}
\end{figure*}

\noindent A scaling ansatz of the form,
\begin{equation}\label{ansatz2}
\text{CDF}(\bar{f})\sim\Delta^{\nu(1+\theta_l)}\;h\,(\bar{f}/\Delta^{\nu}), 
\end{equation}
with $\bar{f}\equiv f^{\prime}/\left<f^{\prime}\right>$ and $\Delta\equiv f_0/\left<f^{\prime}\right>$ fits the data equally well as the scaling form in Eq.~\ref{ansatz}. See fig.~\ref{finite_size}(c). Here, $h(y)$ is the
scaling function such that $\,h(y)\sim y^{1+\theta_l}$ as $y \to \infty$ and it recovers the passive scaling relation $\text{CDF}(\bar{f})\sim \bar{f}^{\,1+\theta_l}$. 
The exponent $\theta_l$ is taken to be $\theta_l=0.15$~\cite{lerner2013low}. In the active regime i.e. in the limit of $y\rightarrow0$, $h(y)\sim(y-y^\prime_c)^\beta$ with $\beta=1.6$ and $y^\prime=0.003$ which is the solid black line in fig.~\ref{finite_size}(c).  

\subsection{Increasing active forcing in smaller increments}
\begin{figure}[h]
    \centering
    \includegraphics[width=0.45\linewidth]{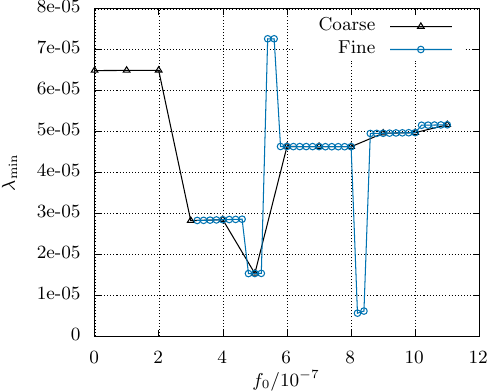}
    \caption{By increasing active force $f_0$ in smaller increments, $\lambda_\text{min}$ does not continuously approach zero. It retains its discrete jumps as a function of $f_0$.}
    \label{ev_fine}
\end{figure}

\end{document}